\documentclass[aps,pre,floatfix,longbibliography]{revtex4-2}
\usepackage{amssymb}
\usepackage{textcomp}
\usepackage{amsmath}
\usepackage{booktabs}
\usepackage{geometry}
\usepackage[figuresright]{rotating}
\usepackage{cmap}
\usepackage[T2A]{fontenc}
\usepackage[utf8x]{inputenc}
\usepackage{hyperref}
\usepackage{xcolor}
\usepackage{graphicx}
\definecolor{color}{HTML}{0000FF}
\hypersetup{linkcolor=color,urlcolor=color,citecolor=color,filecolor=color,colorlinks=true,pdfauthor=author}

\newcommand{\be}{\begin{equation}}
\newcommand{\ee}{\end{equation}}
\newcommand{\bea}{\begin{eqnarray}}
\newcommand{\eea}{\end{eqnarray}}
\newcommand{\p}{\partial}
\newcommand{\vvect}{\mathbf{v}}

\newcommand{\mynabla}{\pmb{\nabla}}

\begin{document}
\title{Influence of fluid flows on electric double layers in evaporating colloidal sessile droplets}
\author{Semen V. Zavarzin$^{1}$}
\author{Andrei L. Kolesnikov$^{2}$}
\author{Yury  A. Budkov$^{1,3}$}
\email{ybudkov@hse.ru}
\author{Lev Yu. Barash$^{3}$}
\email{barash@itp.ac.ru}
\affiliation{$^1$ School of Applied Mathematics, HSE University, 101000 Moscow, Russia}
\affiliation{$^2$ Institut f\"{u}r Nichtklassische Chemie e.V., Permoserstr. 15, 04318 Leipzig, Germany}
\affiliation{$^3$ Landau Institute for Theoretical Physics, 142432 Chernogolovka, Russia}
\begin{abstract}
A model is developed for describing the transport of charged colloidal particles in an evaporating sessile droplet on the electrified metal substrate in the presence of a solvent flow. The model takes into account the electric charge of colloidal particles and small ions produced by electrolytic dissociation of the active groups on the colloidal particles and solvent molecules. We employ a system of self-consistent Poisson and Nernst--Planck equations for electric potential and average concentrations of colloidal particles and ions with the appropriate boundary conditions. The fluid dynamics, temperature distribution and evaporation process are described with the Navier--Stokes equations, equations of heat conduction and vapor diffusion in air, respectively. The developed model is used to carry out a first-principles numerical simulation of charged silica colloidal particle transport in an evaporating aqueous droplet. We find that electric double layers can be destroyed by a sufficiently strong fluid flow.
\end{abstract}
\maketitle
\section{Introduction}
To understand physical mechanisms of pattern formation in the process of sessile droplet drying is a fundamental task
of crucial importance in various applications including biosensors, labs-on-chip, functional coatings, optoelectronics,
biomedical applications, inkjet printing and many others~\cite{tarasevich2019,KolegovBarash2020}.
Although issues of spatial structuring of deposition patterns from drying sessile drops have been extensively studied and the role
of many effects has been clarified~\cite{tarasevich2019,Larson2014,KolegovBarash2020,Han20121534,Sefiane2018,Patil2019,Wang20136048,Brugarolas2013,Zhong201513,Lotito2017217,Tarafdar2018,Patel2018,KolegovBarash2019,Lin2012,Innocenzi2013},
no comprehensive analysis of pattern formation has been done so far, so evaporation of a droplet remains a complex phenomenon
and there are a lot of questions that are unanswered yet. 
A prime example of such a controversial issue is the role of particle--particle, particle--substrate and particle--liquid--gas 
interface interactions in the presence of charged particles and counterions that are caused by electrolytic dissociation.

An influence of the electric double layer (EDL) on the morphology of the deposit has been identified
in the literature in recent years.
In~\cite{Bhardwaj2010,Dugyala2014} the authors altered the pH value of the solution,
which determines the charge on the particles and the interaction forces.
Depending on the pH value, single-ring deposits, a homogeneous spot, or scattered aggregates were obtained.
In~\cite{Morales2013,Crivoi2015,Anyfantakis2015B} charged surfactants were added to alter
the surface charge and electrical potential of suspended particles and hence the EDL force.
The addition of electrolytes to the suspension resulted in a reduction of the Debye length of the EDL
and hence in the suppression of the EDL force~\cite{Kuncicky2008,Nguyen2013,Yan2008}.
The contribution of the van der Waals and EDL forces to the deposition process
is mediated through the rates of particle coagulation in the suspension and particle
adsorption to the substrate~\cite{Yan2008,Homede2018,Homede2020,Zigelman2018}.
The relative role of particle--substrate and particle--particle interactions was investigated
through studying the contact line dynamics of shrinking drops of suspensions~\cite{Moraila2013},
through employing substrates of different wetting properties~\cite{Noguera2015}, and
through modifying the nature of the substrate and particles as well as their size, 
concentrations and surface chemistry~\cite{Lee2017,Bridonneau2020}.
Those experimental results suggest that the contribution of the particle--substrate interaction to the morphology of the deposit
is smaller than the contribution of the particle--particle interactions.

It is known that the deposit structure can be substantially influenced by particle--particle interactions.
This was found experimentally for the case of hydrophobic attraction between particles~\cite{Anyfantakis2017}.
However, various theoretical aspects of this issue have not been carefully studied until now.
The present work focuses on the effects of Coulomb interactions for the case of hydrophilic particles.
In particular, SiO$_2$ particles are known to be hydrophilic~\cite{Williams1974} and so are not affected by hydrophobic attraction.
We note that the hydrophilic repulsion is a surface force that is generally weaker in magnitude as compared to the hydrophobic attraction.

Generally speaking, a negative charge on silica colloidal particles in aqueous sessile droplets can influence their spatial distribution within the droplet volume. In our opinion, in order to understand the role of charge in the transport of silica colloidal particles in the volume of an evaporating aqueous droplet, it is necessary to solve the Nernst--Planck equation for the average concentrations of colloidal particles and counterions taking into account the Marangoni flow. The Nernst--Planck equation is the basic equation, describing the diffusion of the ions in dilute electrolyte solutions under various conditions, such as electroosmosis \cite{park2007comparison}, ionic current through the channels in the lipid membranes of living cells \cite{kurnikova1999lattice}, charging of the electric double layers in batteries and capacitors \cite{wang2013simulations,lopez2019ionic}. There are several papers (see recent ones \cite{warren2020non,warren2019diffusiophoresis} and review \cite{shin2020diffusiophoretic} with the references therein), where the authors studied the effect of diffusiophoresis on the distribution of the charged colloidal particles suspended in electrolyte solutions by solving the Nernst--Planck equation without taking account solvent flows. One should also mention paper \cite{rivas2018mesoscopic}, where the authors proposed an approach based on the lattice Boltzmann equation and Nernst--Planck equations to describe the kinetics of the average concentrations of colloidal particles and ions in a spherical liquid droplet. However, to the best of our knowledge, joint effects of the capillary and Marangoni flows of the solvent on the spatial distribution of charged colloidal particles in salt-free aqueous media have not been addressed properly in the literature up to now. 

The aim of this work is to fill this gap and to develop a sufficiently general theoretical approach to these issues, based on a self-consistent Poisson and Nernst--Planck system of equations and without using the DLVO approximation. Such an approach allows one to perform numerical simulations of transport of charged particles in droplets for a variety of problems. Here, we use our approach to describe the influence of fluid flows on electric double layers in evaporating aqueous sessile droplets containing silicone oxide particles.
We consider two types of naturally occurring liquid flows in a droplet: a thermocapillary convective Marangoni flow, which appears due to the temperature dependence of liquid surface tension~\cite{HuLarson2005,Barash2009}, and a capillary outward flow (aka compensatory flow) that originates from a nonuniform evaporation profile along the droplet surface and is known to result in a coffee-ring effect~\cite{Deegan1997,Deegan2000}. 
The thermocapillary convection usually dominates the capillary flow at comparatively large contact angles, $\theta$, while the latter becomes dominating only at the final stage of droplet evaporation, which means that the flows are weak at sufficiently small $\theta$~\cite{KolegovBarash2020}.
Our basic result is that a strong fluid flow, such as the naturally occurring Marangoni convective flow in the system under consideration, destroys the electric double layer.

In our approach to simulations of spatial distributions of temperature and fluid velocity in an axially symmetrical droplet, we employ the results of Refs.~\cite{Barash2009,HuLarson2002}. 
In particular, the basic relations for heat conduction and vapor diffusion in air are summarized in Appendix~\ref{appendix-temperature}. To obtain the Marangoni flow, we use the equations and boundary conditions which are summarized in Appendix~\ref{appendix-velocity}. The capillary flow is obtained with the boundary conditions developed in Appendix~\ref{appendix-capillary}.
The colloidal droplet physical parameters used in the work are represented in Table~\ref{tab:parameter}. 

The main part of the paper is organized as follows. In Sec.~\ref{sec:equations} we present basic equations and boundary conditions describing the transport of charged particles in droplets. In Sec.~\ref{sec:results} we apply our approach to simulations of dynamics of electric potential and colloidal particle concentration under the influence of fluid flows. Sec.~\ref{sec:conclusions} contains the discussion.

\begin{table*}

		\caption{Parameter values used in the paper}\label{tab:parameter}
		\begin{tabular}{p{3cm} p{5.5cm} p{6cm} }
			\hline \hline
			&& \\
	    	Drop parameters & Contact line radius & $R = 5 \times 10^{-3}~m$ \\ 
			& Contact angle & $\theta=30^{0}$  \\ 
			& &\\
			Substrate parameters & Thickness & $H = 5 \times 10^{-3}~m$ \\
			&Temperature & $T = 298.15 K$ \\
			&&\\
			Water characteristic & Density & $n=977~kg/m^{3}$\\
			&Heat capacity & $c_p=4181.6~J/(kg~K)$\\
			&Thermal conductivity& $k=0.6069~W/(m~K)$\\
			&Thermal diffusivity& $\kappa=k/ (\rho c_p) = 1.46 \times 10^{-6}~m^{2}/s$\\
			&Dynamic viscosity& $\eta=8.9 \times 10^{-4}~kg  / (m~s)$\\
			&Latent heat of evaporation& $L = 2.44 \times 10^{6}~J  / kg$ \\
			&Diffusion coefficient of aqueous vapour in air& $D=2.42 \times 10^{-5}~m^{2}/s$\\
			&Saturated vapor density & $u_s=2.35 \times 10^{-2}~kg  /  m^{3}$ \\
			&Temperature derivative of surface tension& $\sigma ' = \partial{\sigma}/\partial{T}=-1.46 \times 10^{-4}~kg/(s^{2}~K)$\\
			&Dielectric constant& $\varepsilon_{s}=80$\\
			&Hydrodynamic radius of counterions ($H_{3}O^{+}$) & $R_+ = 3 \times 10^{-10}~m$\\
			&Hydrodynamic radius of coions ($OH^{-}$) & $R_- = 1.53 \times 10^{-10}~m$\\
			&&\\
			
			Silica particles characteristics& Hydrodynamic radius of colloidal particles (SiO$_2$) & $R_c = 3.5 \times 10^{-7}~m$\\ 
			&Average concentration of colloidal particles (SiO$_2$)& $n_{c,0}=1.856 \times 10^{13}~m^{-3}$\\
			& Charge& $q_c=-1000e$\\
			&Dielectric constant & $\varepsilon_{c}=4$\\
						
			\hline \hline
		
		\end{tabular}

\end{table*}

\section{Basic equations and boundary conditions}
\label{sec:equations}
\begin{figure}
	\centering
	\includegraphics[width=0.5\linewidth]{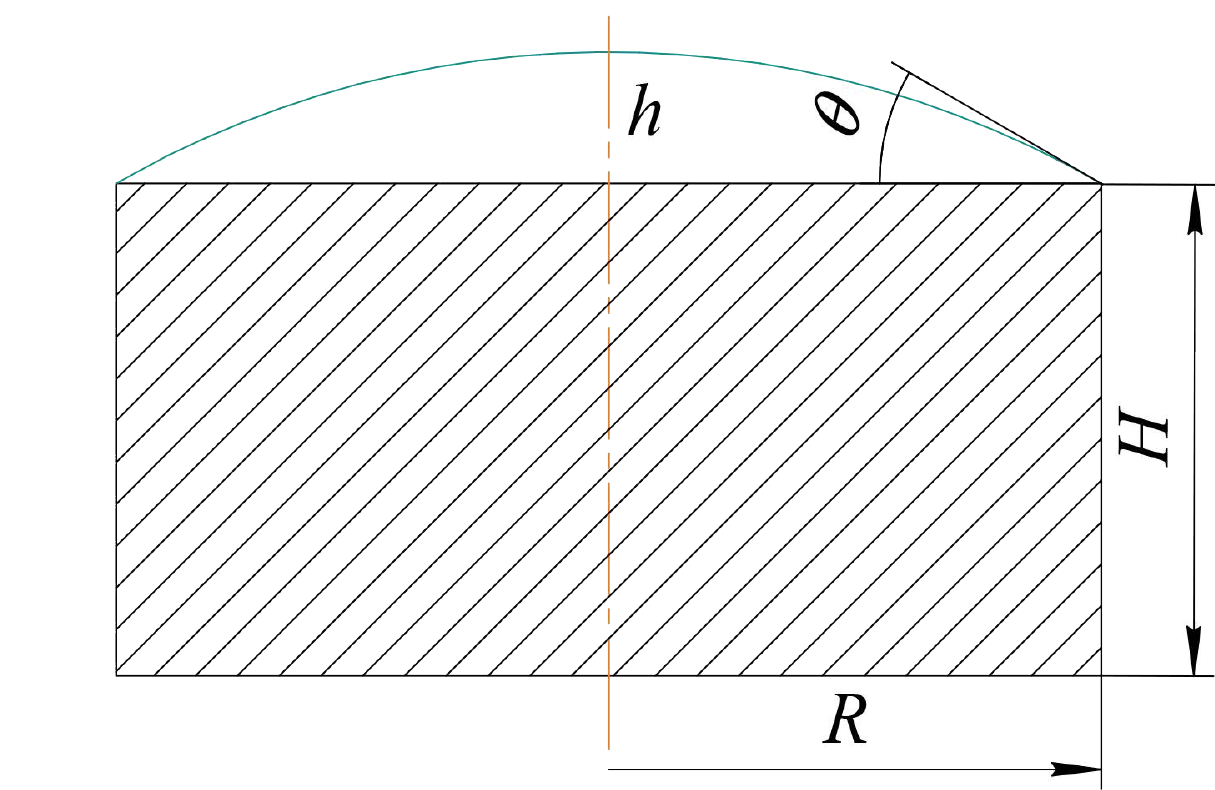}
	\caption{Sessile droplet on a metal cylindrical substrate.}
	\label{fig:drop}
\end{figure}

Let us consider a liquid droplet that has the form of a spherical segment with a contact angle $\theta$ and a radius of the base $R$ that is placed on a cylindrical metal substrate (electrode) with a height $H$ and the same radius $R$ under the fixed applied voltage, $\Phi$ (see Fig.~\ref{fig:drop}). We assume that $N_{c}$ colloidal particles with a charge $q_{c}$ and $N_{+}+N_{-}$ small ions with a charge $q_{\pm}=\pm e$ ($e$ is the elementary charge) are suspended within the droplet volume. Small ions appear in the solution due to the electrolytic dissociation of active groups on the surfaces of colloidal particles and water molecules. The system as a whole is electrically neutral, so that the condition $q_{c}N_{c}+q_{+}N_{+}+q_{-}N_{-}=0$ is fulfilled. We also assume that the electrode is connected with the thermostat with a fixed temperature $T_0$. In what follows, we formulate the basic equations in the general form which is applicable to any geometry of a system. Nevertheless, in Appendix~\ref{appendix-axial} for convenience we rewrite them in the cylindrical coordinates taking into account the axial symmetry.

The electrostatic potential, $\phi(\mathbf{r},t)$, inside the droplet can be obtained from the solution of the Poisson equation
\begin{equation}
\label{Poisson}
\nabla\cdot(\varepsilon(\mathbf{r},t)\nabla\phi(\mathbf{r},t))= -\frac{\rho(\mathbf{r},t)}{\varepsilon_0},
\end{equation}
where
\begin{equation}
\varepsilon(\mathbf{r},t)=\varepsilon_s +\frac{3v(\varepsilon_c-\varepsilon_s)\varepsilon_s}{\varepsilon_c+2\varepsilon_s}n_c(\mathbf{r},t)
\end{equation}
is the relative dielectric permittivity of colloidal suspension \cite{landau2013electrodynamics}, where we have introduced the dielectric permittivities $\varepsilon_s$, $\varepsilon_c$ of solvent and colloidal particles, respectively, and volume $v=4\pi R_c^3/3$ of the spherical colloidal particles with the hydrodynamic radius $R_c$; $\varepsilon_{0}$ is the permittivity of vacuum; $n_{c}(\mathbf{r},t)$ is the local concentration of colloidal particles (see below); $\rho(\mathbf{r},t)$ is the local charge density which can be written as a sum of two terms
\begin{equation}
\rho(\mathbf{r},t)=\rho_{ion}(\mathbf{r},t)+\rho_c(\mathbf{r},t),
\end{equation}
where
\begin{equation}
\rho_c(\mathbf{r},t)=q_c n_{c}(\mathbf{r},t)
\end{equation}
is the charge density of the colloidal particles with a charge $q_c$ and the aforementioned local concentration $n_c(\mathbf{r},t)$, which for the case of a sufficiently dilute colloidal solution can be obtained by solving the Nernst--Planck equation taking into account the solvent motion
\begin{equation}
\label{Fokker}
\frac{\partial{n_c}}{\partial{t}}=\frac{q_c}{\gamma_c}\nabla\cdot\left(n_c\nabla{\phi}\right)+\nabla\cdot (D_c \nabla{n_c})-\nabla\cdot (n_c\mathbf{v}),
\end{equation}
where $\gamma_c=6\pi\eta R_c$, $R_c$ is the hydrodynamic radius of the colloidal particles, $\eta$ is the solvent viscosity, $D_c(\mathbf{r},t)=k_{B}T(\mathbf{r},t)/\gamma_c$ is the diffusion coefficient of the colloidal particles; $k_B$ is the Boltzmann constant, $T(\mathbf{r},t)$ is the local temperature, which can be obtained from the solution of the heat transfer equation in the droplet with appropriate boundary conditions (see Appendix~\ref{appendix-temperature}); the last term in the right hand side of eq. (\ref{Fokker}) describes the convection term occurring due to the liquid motion with the field of velocity $\mathbf{v}(\mathbf{r},t)$, which must be obtained from the solution of hydrodynamics equations with the appropriate boundary conditions (see Appendix~\ref{appendix-velocity}).

The total droplet drying time can be estimated as $t_f \approx 0.2 n R h_0/ (Du_s) \approx 38$~min.~\cite{Larson2014}, so it is much longer than all other characteristic time scales of the problem. Hence, it is not necessary to account for the temporal evolution of the droplet profile, and the contact angle value $\theta$ can be fixed during the simulation.

We also neglect the influence of the electrophoresis of the colloidal particles and counterions on the hydrodynamics of liquids. The latter can be justified by sufficiently small volume fraction of the colloidal particles and macroscopic size of the droplet. We note that the effects of ion electrophoresis on liquid velocity distribution can be relevant for the solvent electroosmosis in electrolyte solutions confined in nanocapillaries under external electrostatic field \cite{maduar2015electrohydrodynamics,park2007comparison,squires2005microfluidics}. The local charge density of the ions is
\begin{equation}
\rho_{ion}(\mathbf{r},t)=q_{+} n_{+}(\mathbf{r},t)+q_{-} n_{-}(\mathbf{r},t),
\end{equation}
where the ionic concentrations, $n_{\pm}(\mathbf{r},t)$, can be obtained from the solution of the similar Nernst--Planck equation
\begin{equation}
\label{Fokker_2}
\frac{\partial{n_i}}{\partial{t}}=\frac{q_i}{\gamma_i}\nabla\cdot\left(n_i\nabla{\phi}\right)+\nabla\cdot(D_i \nabla{n_i})-\nabla\cdot(n_i\mathbf{v}),~i=\pm
\end{equation}
where $\gamma_i=6\pi \eta R_i$ with the hydrodynamic radii of the ions, $R_i$, and $D_{i}(\mathbf{r},t)=k_{B}T(\mathbf{r},t)/\gamma_i$ is their diffusion coefficient. In order to solve the systems of equations (\ref{Poisson}), (\ref{Fokker}), and (\ref{Fokker_2}), we have to formulate the appropriate boundary conditions. Outside the droplet and metal substrate the Laplace equation for the electrostatic potential must be solved as
\begin{equation}
\Delta \phi =0,
\end{equation}
so that the boundary conditions take the following form
\begin{equation}
\label{Bound}
\phi_{d}^{(sub)}=\Phi,~\phi_{sub}=\Phi,~\varepsilon \frac{\partial{\phi}_{d}^{(+)}}{\partial{\mathbf{n}}}=\frac{\partial{\phi}_{d}^{(-)}}{\partial{\mathbf{n}}},
~\phi_{d}^{(-)}=\phi_{d}^{(+)},~\phi\rvert_{|\mathbf{r}|\to \infty}=0,
\end{equation}
where $\phi_{d}^{(sub)}$ is the potential on the droplet-substrate interface and $\phi_{sub}$ is the substrate potential which has a fixed value, $\Phi$; $\phi_{d}^{(\pm)}$ are the potentials on the inner ($+$) and outer ($-$) droplet surfaces; the $\partial{}/\partial{\mathbf{n}}$ symbol denotes the normal derivative. We assume that the metal substrate and the droplet surface are impermeable for colloidal particles and counterions, which leads to the following boundary conditions for the local concentrations $n_{c,\pm}(\mathbf{r},t)$
\begin{equation}
\label{Bound_2}
\frac{\partial{n_{c,\pm}}}{\partial{\mathbf{n}}}+\frac{q_{c,\pm}n_{c,\pm}}{k_{B}T}\frac{\partial{\phi}_{d}^{(+)}}{\partial{\mathbf{n}}}=0,
~\frac{\partial{n_{c,\pm}}}{\partial{\mathbf{n}}}+\frac{q_{c,\pm}n_{c,\pm}}{k_{B}T}\frac{\partial{\phi}_{sub}^{(+)}}{\partial{\mathbf{n}}}=0
\end{equation}
where we take into account that the normal component of the velocity on the droplet surface and on the substrate is equal to zero. The initial conditions for the concentrations of colloidal particles and counterions are fixed functions
\begin{equation}
n_{c}(\mathbf{r},0)=n_{c}(\mathbf{r}),~n_{\pm}(\mathbf{r},0)=n_{\pm}(\mathbf{r}),
\end{equation}
which must satisfy the global electroneutrality condition, i.e. 
\begin{equation}
\label{neutrality}
\int\limits_{V}d\mathbf{r}(q_{+}n_{+}(\mathbf{r})+q_{-}n_{-}(\mathbf{r})+q_{c}n_{c}(\mathbf{r}))=0.
\end{equation}
In particular, we can assume the uniform initial distributions, $n_{c,\pm}(\mathbf{r})=N_{c,\pm}/V$, of colloidal particles and ions inside the droplet volume. Conditions (\ref{neutrality}) and (\ref{Bound_2}) together with Eqs.~(\ref{Fokker})--(\ref{Fokker_2}) automatically result in the global electrical neutrality condition at $t>0$.

\section{Results and discussion}
\label{sec:results}
Using the FlexPDE 7.18 software, we simulated transport of charged colloidal particles in an evaporating aqueous droplet having the shape of a spherical segment with the base radius, $R = 5~mm$, and contact angle, $\theta=30^{0}$ (Fig.~\ref{fig:drop}) and placed on a thermostatically controlled electrified metal substrate. In accordance with Sec.~\ref{sec:equations}, we numerically solved the system of coupled equations: Nernst--Planck equations for the average concentrations of colloidal particles and counterions, the Poisson/Laplace equation for the electrostatic potential inside/outside the droplet volume, the heat transfer equation for the temperature inside the droplet, and the Navier--Stockes equations for the liquid velocity distribution. The droplet contains an aqueous suspension of SiO$_{2}$ colloidal particles. The colloidal particles (total number $N_{c}=10^6$) carry the charge $q_{c}=-1000e$. We also assume that each colloidal particle possesses the hydrodynamic radius $R_{c}= 0.35~{\mu}m$. The dielectric permittivity of silica colloidal particles is $\varepsilon_c=4$ (SiO$_2$), while the dielectric permittivity of water is assumed to be $\varepsilon_{s}=80$. It is assumed that the counterions ($H_{3}O^{+}$ in this case) and coions ($OH^{-}$) have the hydrodynamic radii $R_{+}\approx 0.3~nm$ and $R_{-}\approx 0.153~nm$, respectively and their charges are assumed to be $q_{\pm}=\pm e$. The metal substrate has the shape of a cylinder with a radius $R$ and height $H=R$. As we have mentioned above, the substrate is connected with the thermostat, so that its temperature, $T_0=298.15~K$, is fixed. The applied voltage, $\Phi$, on the substrate is in the interval of $-0.5~V$ -- $1.0~V$ and is responsible for the electrochemical stability of water.

\begin{figure}
	\centering
	\includegraphics[width=0.5\linewidth]{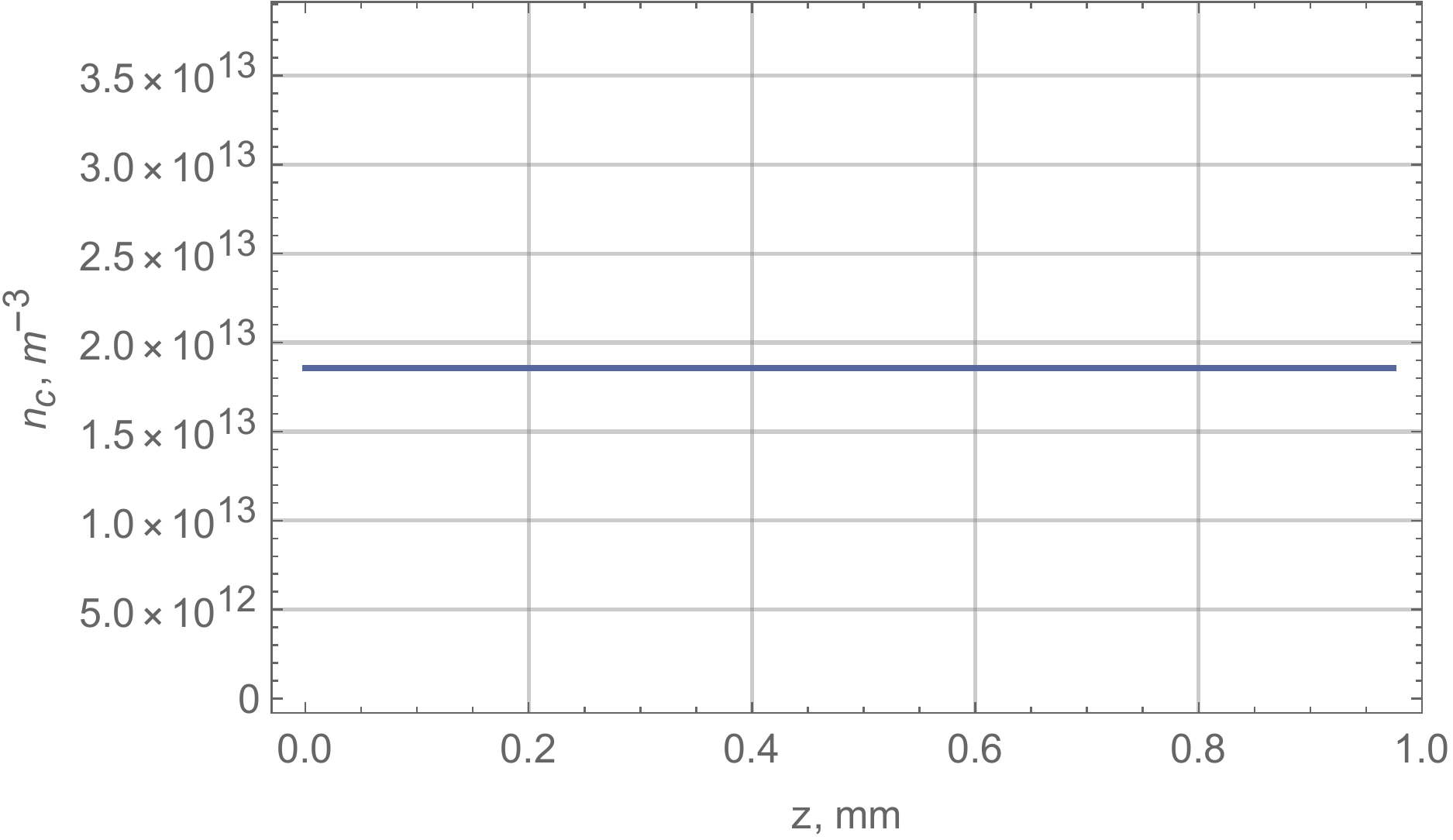}
	\caption{Typical colloidal particle concentration curve as a function of $z$ at $r=0.5R$ at the zero voltage. The flows are switched on.}
	\label{fig:ncnoel}
\end{figure}

\begin{figure}
	\centering
	\includegraphics[width=0.5\linewidth]{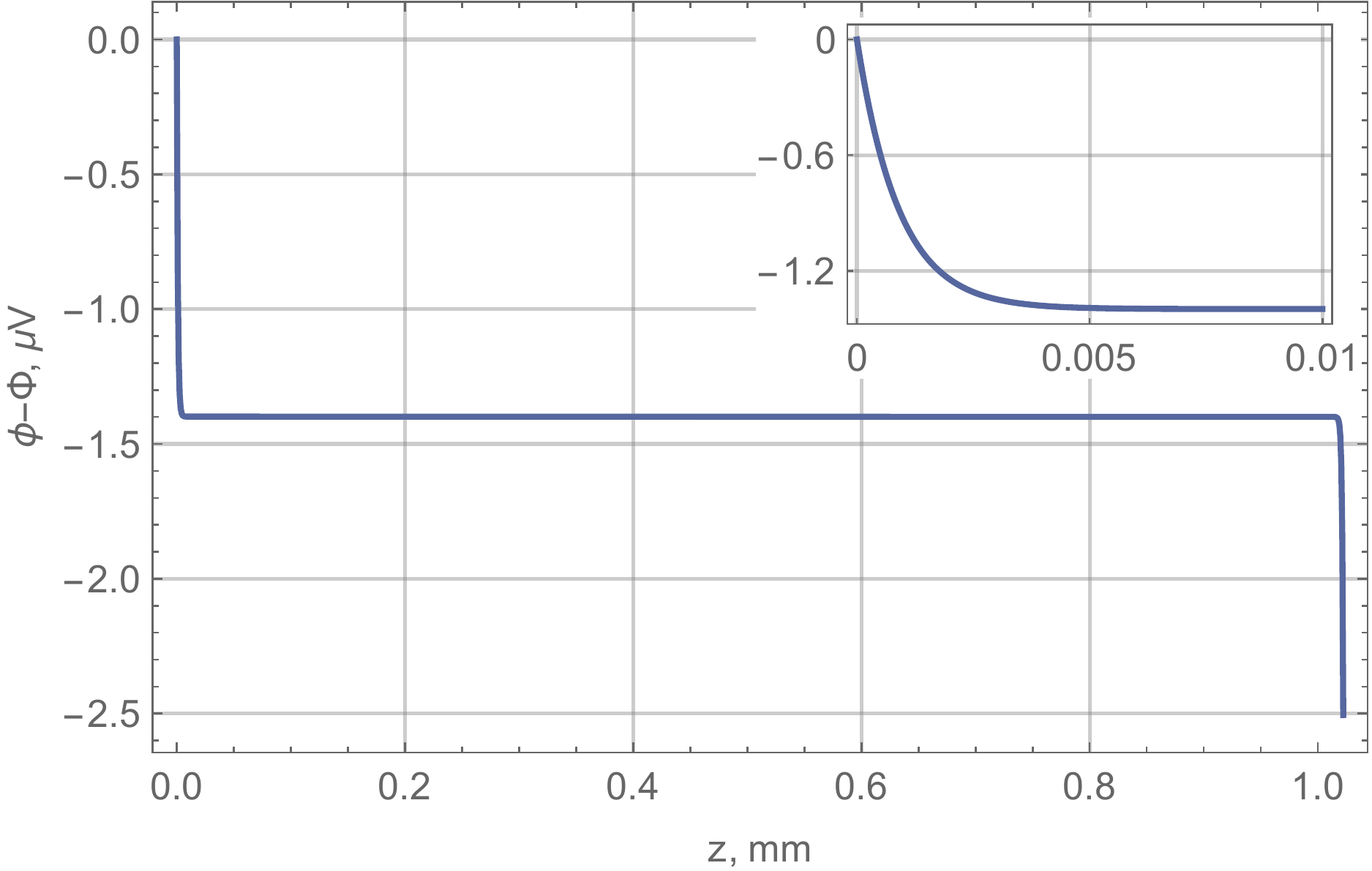}
	\caption{Potential drop as a function of $z$ at $r = 0.5R$ from the droplet center at voltage $\Phi=0.5~V$. The flows are switched off.}
	\label{fig:potential}
\end{figure}

\begin{figure*}
	\centering
	\includegraphics[width=0.4\textwidth]{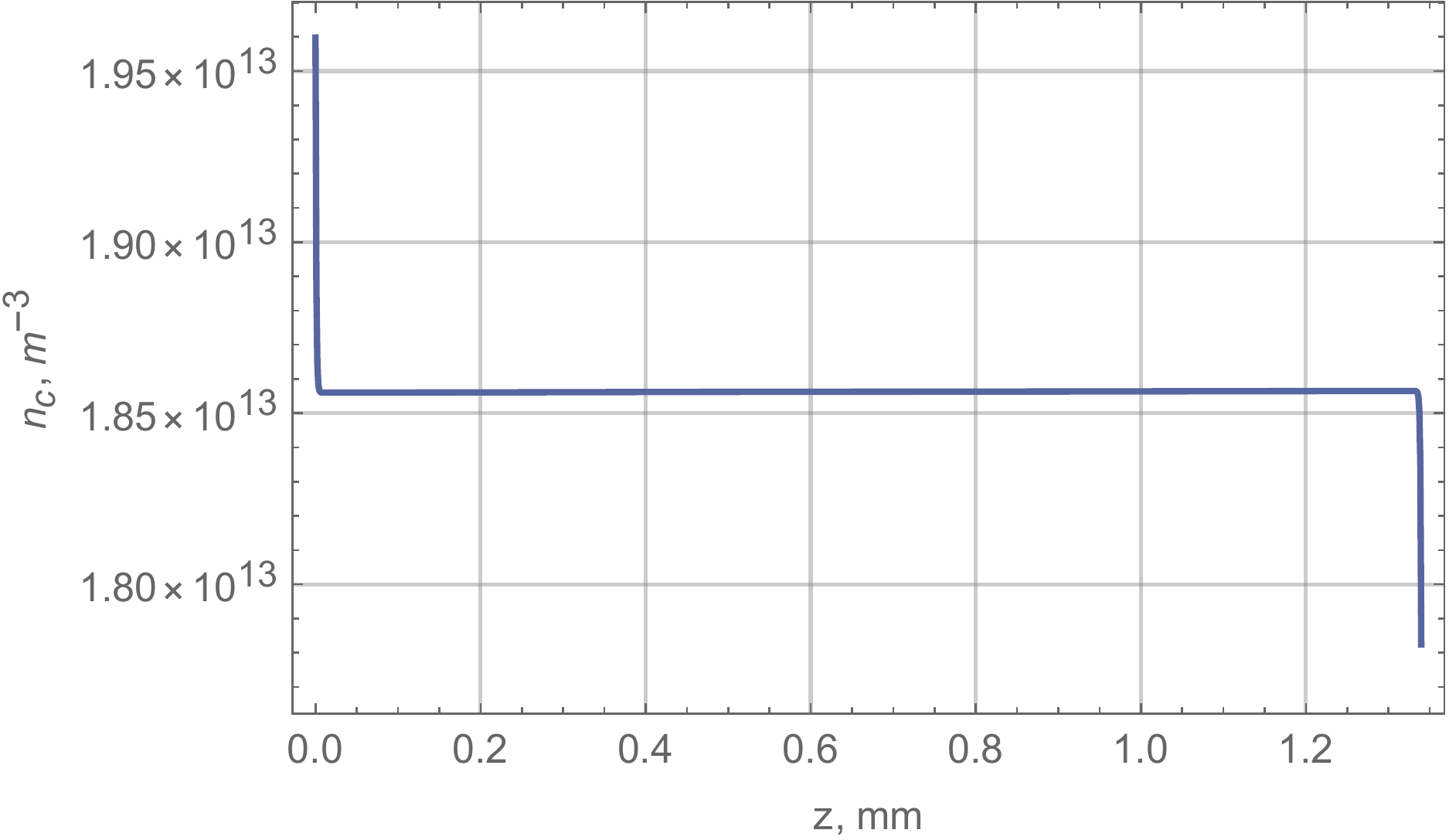}
	\includegraphics[width=0.4\textwidth]{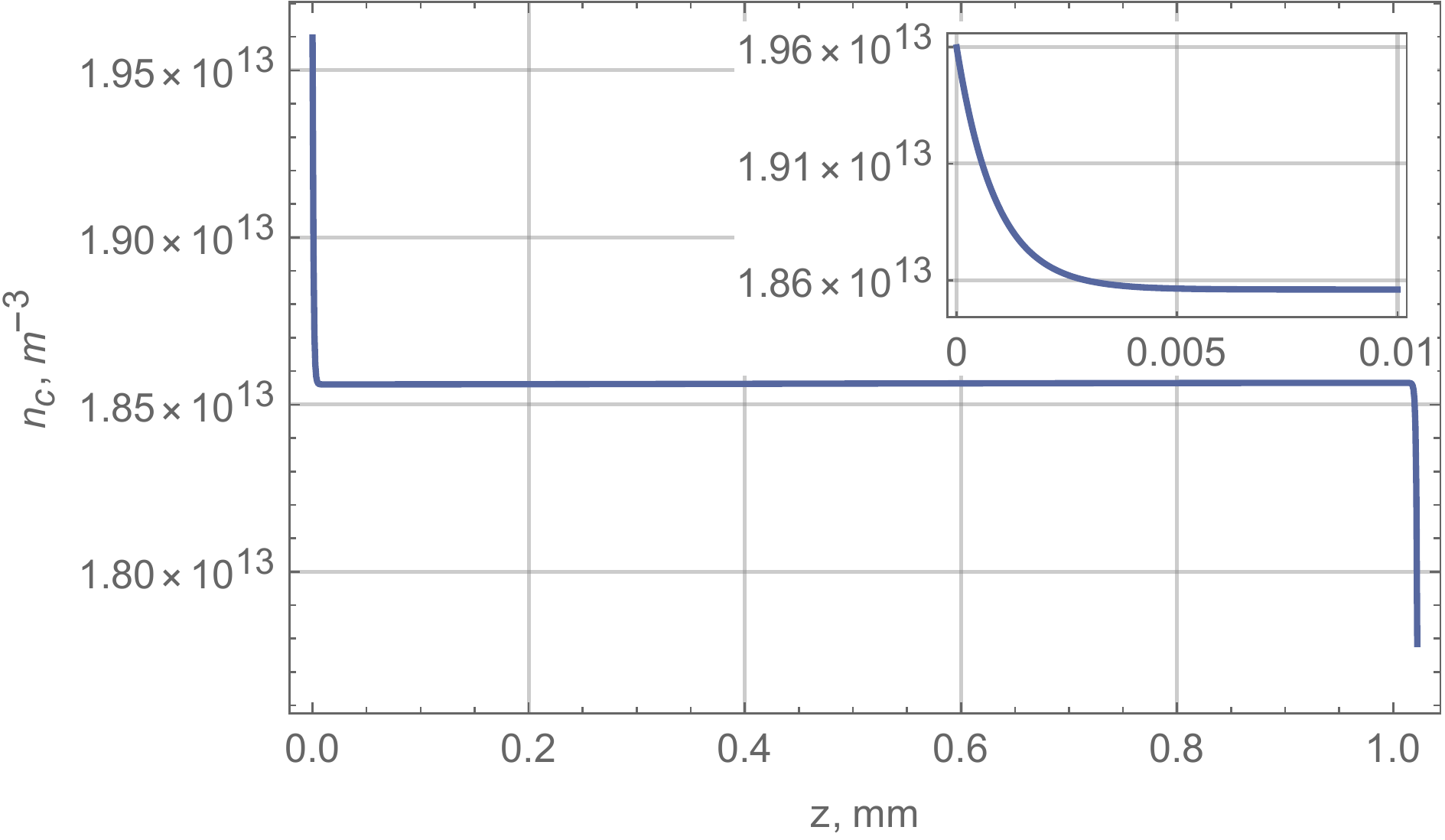}
	\caption{Colloidal particle concentration curve as a function of $z$ at $r = 0$ (on the left) and $r = 0.5R$ (on the right) from the droplet center at the voltage $\Phi=0.5~V$. The flows are switched off.}
	\label{fig:ncpnf}
\end{figure*}

\begin{figure*}
	\centering
	\includegraphics[width=0.4\textwidth]{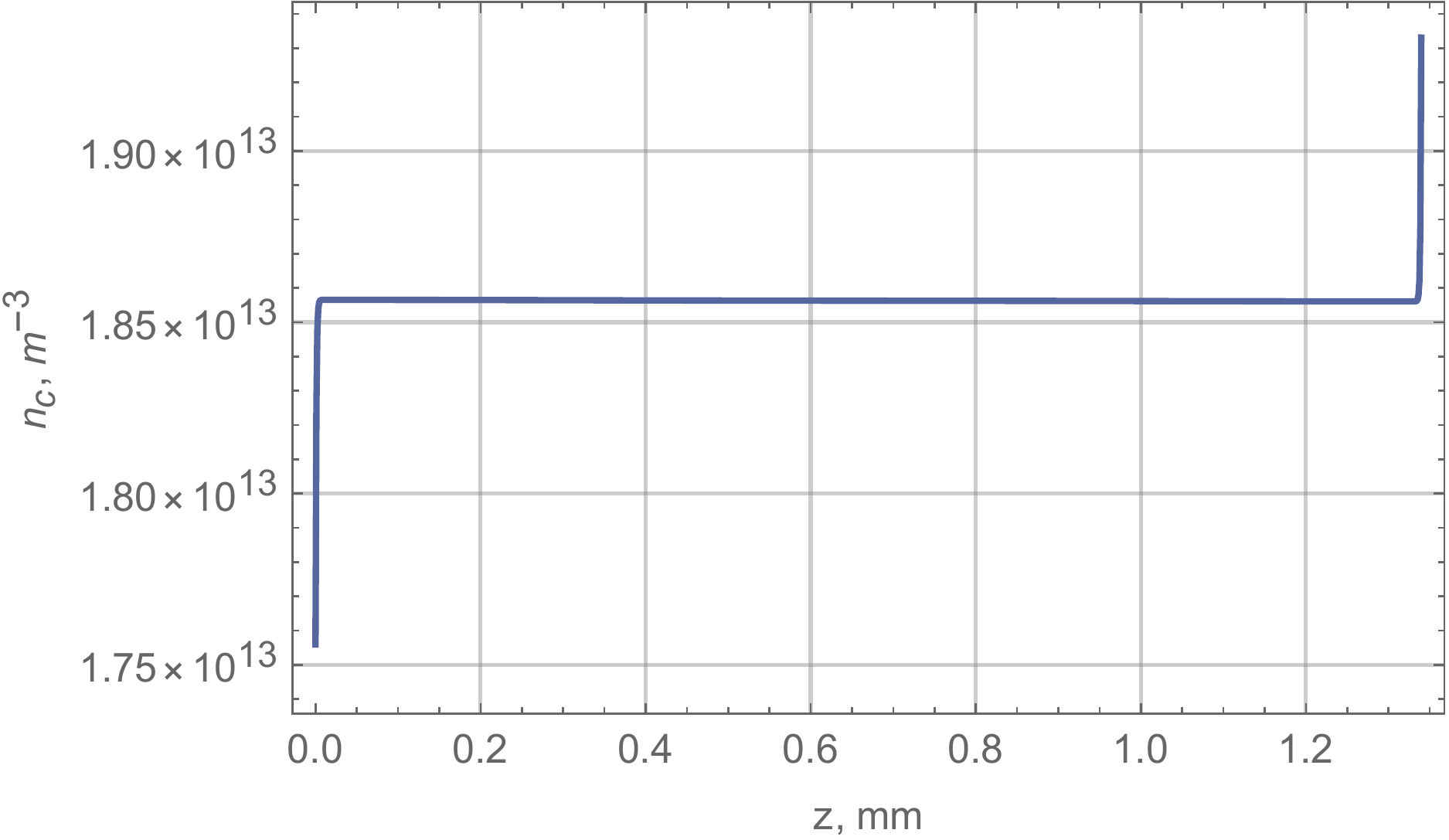}
	\includegraphics[width=0.4\textwidth]{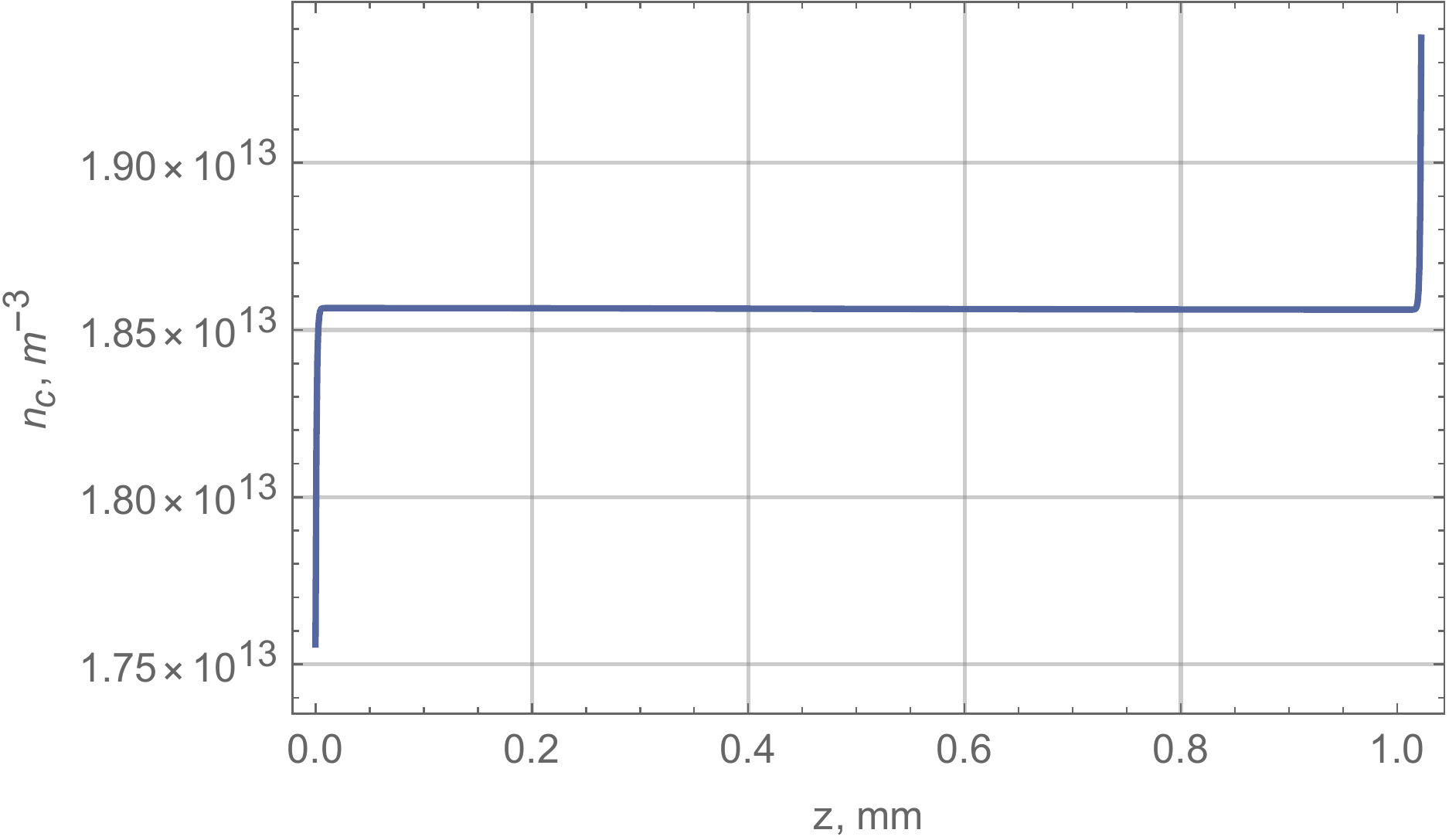}
	\caption{Colloidal particle concentration curve as a function of $z$ at $r = 0$ (on the left) and $r = 0.5R$ (on the right) from the droplet center at the voltage $\Phi=-0.5~V$. The flows are switched off.}
	\label{fig:ncnnf}
\end{figure*}

Firstly, we considered the case of a zero potential drop on the metal substrate. In the case of zero voltage, the colloidal particles and the counterions were distributed quite uniformly with concentrations $n_{c,0}=N_{c}/V=1.856 \times 10^{13}~m^{-3}$, $n_{+,0}=N_{+}/V=6.022\times 10^{19}~m^{-3}$, $n_{-,0}=N_{-}/V=6.02\times 10^{19}~m^{-3}$ (see Fig.~\ref{fig:ncnoel}). The concentrations of the ions were calculated from the condition pH$=7$. The spatial dependence of the diffusion coefficient through the temperature and the fluid motion did not influence the colloidal particle distribution. However, the positive voltage applied to the substrate made the colloidal particles accumulate at the liquid--substrate interface during the first several seconds of the simulation. The latter is related to the formation of an EDL at the liquid/metal substrate interface. At the same time, the counterions at the liquid--gas interface also formed an EDL. The negative voltage on the substrate also resulted in EDL formation. However, in this case, the counterions were accumulated at the liquid--substrate interface, while the colloidal particles -- at the droplet surface. If there were only capillary flows and no Marangoni flow, we observed the formation of EDLs with a constant thickness (see Figs.~\ref{fig:potential},~\ref{fig:ncpnf},~\ref{fig:ncnnf}), which is approximately equal to the Debye radius
\begin{equation}
r_{D}=\left(\frac{q_{c}^2N_{c}+q_{+}^2N_{+}+q_{-}^2N_{-}}{\varepsilon_{s}\varepsilon_{0}k_{B}TV}\right)^{-1/2}\approx 0.9~{\mu}m.
\end{equation}
As it follows from the classical EDL theory, if the local colloidal particle concentration on the substrate, $n_{c,s}\approx1.962\times 10^{13}~m^{-3}$, and concentration in the bulk solution, $n_{c,b}\approx1.856\times 10^{13}~m^{-3}$, the droplet should satisfy the relation $n_{c,s}/n_{c,b}\approx\exp\left[-q_{c}\delta\phi/(k_{B}T)\right]$, where $\delta\phi$ is the difference between the potentials of the substrate and the bulk solution. Using the calculated value $\delta\phi\approx  1.43~\mu V$ (see Fig.~\ref{fig:potential}), we obtain $\exp\left[-q_{c}\delta\phi/(k_{B}T)\right]\approx 1.057$, which is equal to the ratio $n_{c,s}/n_{c,b}\approx 1.057$, as can be seen from the inset in Fig. \ref{fig:ncpnf}. We note that we can safely neglect the electrowetting effect in our simulations, because the EDL formation in our case leads to a very small decrease in the contact angle. Indeed, in accordance with the Lippmann equation~\cite{squires2005microfluidics} the increase in the cosine of the contact angle can be estimated as $\Delta (\cos{\theta})=C\Phi^2/2\sigma$, where $\sigma\simeq 73\times 10^{-3}~N m$ is the liquid--vapor surface tension of water and $C\sim \varepsilon_s\varepsilon_{0}/r_{D}$ is the electric capacitance of the EDL. Thus, at $\Phi=1~V$ we have a negligible value of $\delta (\cos{\theta}) \simeq 5.4\times 10^{-3}$. We also note that the influence of colloid particles on the dielectric permittivity (dielectric mismatch) is negligible in the extremely dilute colloid solution considered here.

\begin{figure*}
	\centering
	\includegraphics[width=0.4\textwidth]{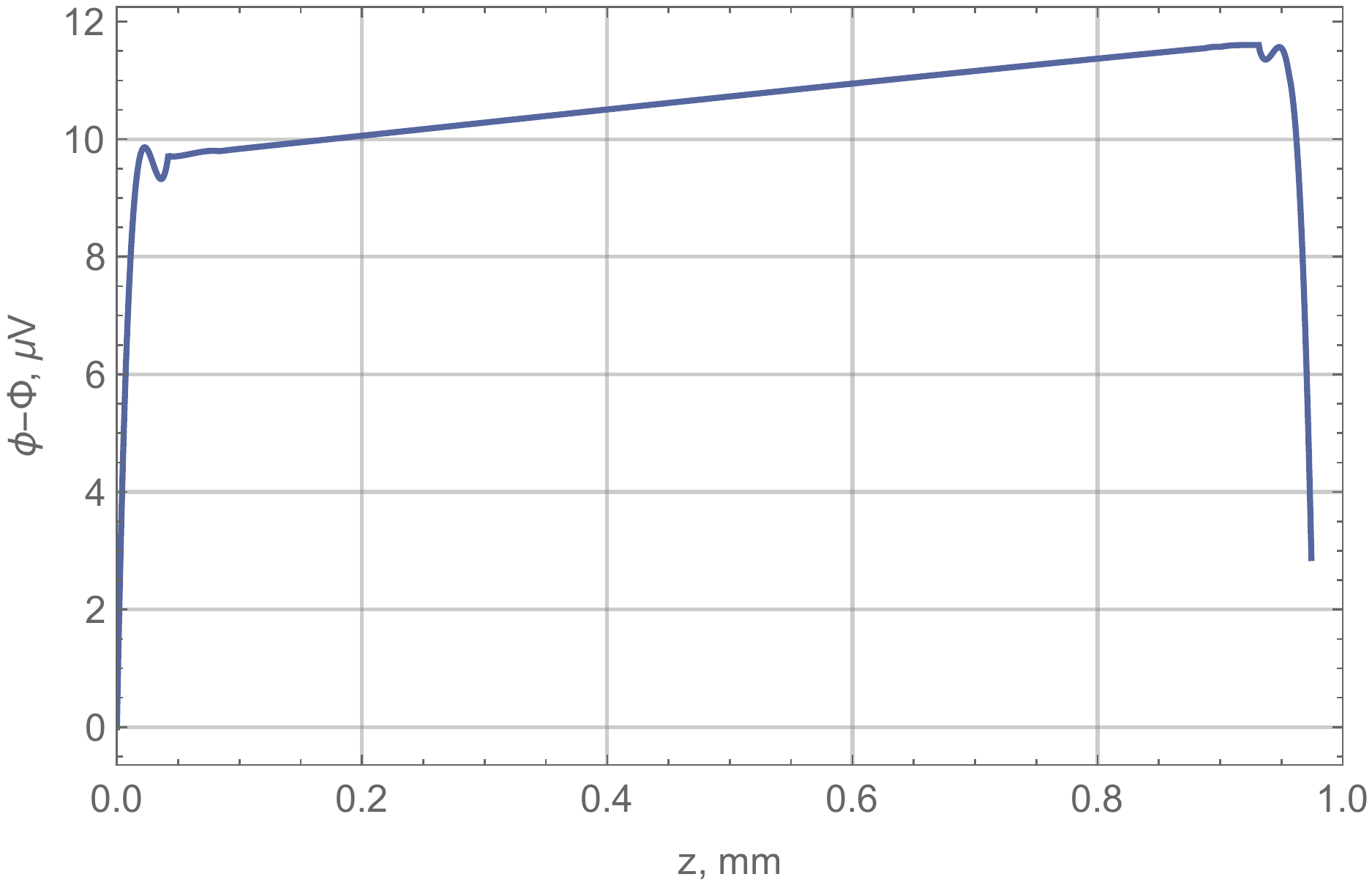}
	\includegraphics[width=0.4\textwidth]{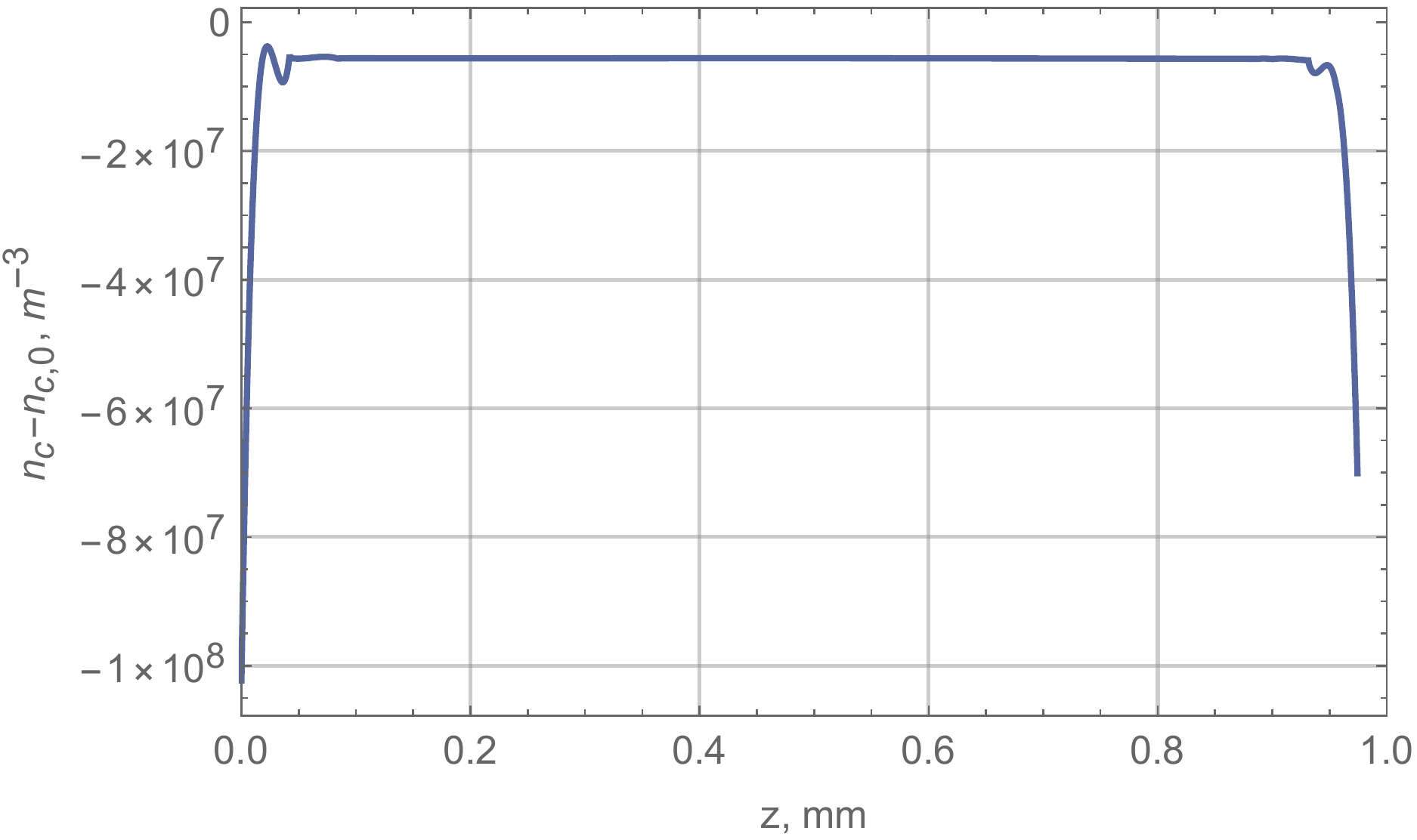}
	\caption{Potential drop (on the left) and colloidal particle concentration curve as a function of $z$ at $r = 0.5R$ (on the right) at voltage 0.5~V. The flows are switched on.}
	\label{fig:ncmc}
\end{figure*}

\begin{figure}
	\centering
	\includegraphics[width=0.5\linewidth]{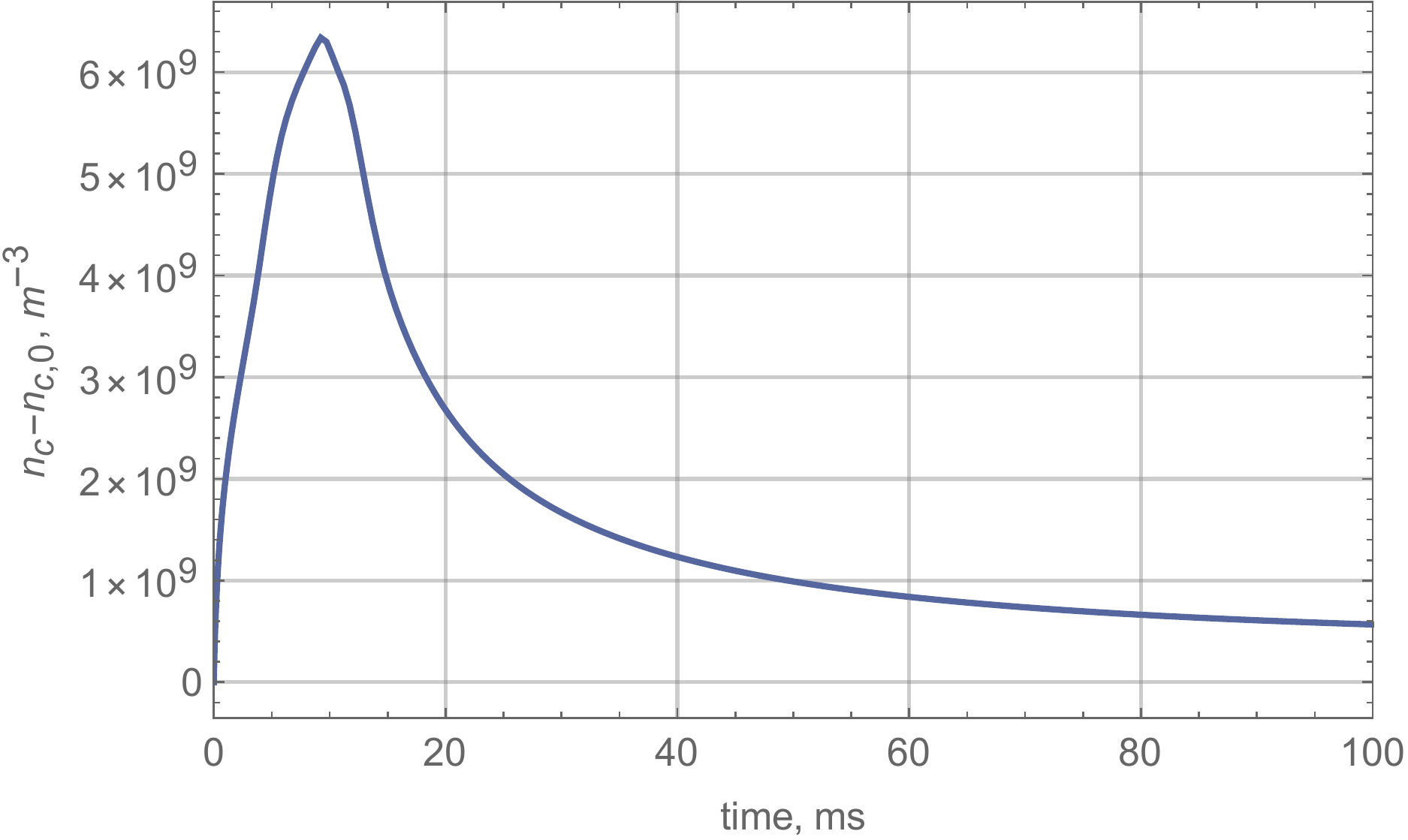}
	\caption{A spatially averaged concentration of colloidal particles on the substrate surface as a function of time. Voltage $\Phi=0.5~V$. The flows are switched on.}
	\label{fig:ncbot}
\end{figure}

However, in the case when the Marangoni flow was switched on, after a sufficiently small time (about $100~ms$) from the simulation start, we observed the destruction of the EDLs due to the fluid mixing. The colloidal particles and counterions were distributed more or less uniformly in the droplet volume (see Fig.~\ref{fig:ncmc}). This is natural if the adhesion to the substrate and capillary attraction between the particles are not taken into account~\cite{KolegovBarash2019,Muzaiqer2021,budkov2013surface}. Nevertheless, Fig.~\ref{fig:ncmc} shows a small lack of colloidal particles near the substrate and free boundary of droplet in comparison with the bulk solution. This effect is a direct consequence of the boundary conditions (\ref{Bound_2}): the normal derivative of the colloidal particle concentration has the same sign as the normal derivative of the electrostatic potential (see the left panel of Fig. \ref{fig:ncmc}). The EDL disruption is also shown in Fig.\ref{fig:ncbot}, where we plotted a time dependence of the difference between the concentration of the colloid particles averaged over the substrate area and their mean volume concentration. As is seen, over a rather short time ($\approx 15~ms$), the difference monotonically increases, then the value reaches a maximum and starts to monotonically decrease.
We also obtain similar results for various values of  $n_{c,0}$ with the total number of colloidal particles $N_c$ up to $10^9$.
We note that the small potential difference between the substrate and the droplet free boundary is the result of the steady ionic currents driven by the Marangoni flow in the droplet volume. Thus, we conclude that in the presence of a Marangoni flow, the electrostatic interactions do not play an important role in the resulting colloidal particles distribution. This conclusion is in qualitative agreement with the results of Ref.~\cite{Bridonneau2020}. Qualitatively analyzing their experimental data for a similar system, the authors of Ref.~\cite{Bridonneau2020} demonstrated, within the DLVO theory, the irrelevance of electrostatic colloidal-substrate interactions in forming colloidal particle distribution inside the droplet. Our study reveals the possible mechanism of such phenomenon. Finally, we note that to avoid EDL disappearance, it is necessary to decrease the Marangoni number by three orders of magnitude.

\section{Conclusions}
\label{sec:conclusions}
We have developed a model for describing transport of charged colloidal silica particles in an evaporating aqueous sessile droplet on an electrified metal substrate in the presence of capillary and Marangoni flows in the solvent. We have formulated a system of self-consistent Poisson and Nernst--Planck equations for electric potential and average concentrations of colloidal particles and counterions with the appropriate boundary conditions. Based on the formulated model, we performed a first-principles numerical simulation of the transport of charged silica particles. We have found that the electric double layers formed on the electrified substrate and the surface of the droplet can be destroyed by a sufficiently strong fluid flow. For the particular system considered, where the capillary fluid flows are much weaker than the Marangoni flows, the electric double layer can be destroyed by the Marangoni flows and not by the capillary flow. Regarding the problem of pattern formation from drying drops, our result implies that a strong fluid flow should alter the rates of particle coagulation in the suspension and particle adsorption to the solid substrate.

\section*{Acknowledgements}
The work of Y.A.B. and L.Y.B was supported by grant No. 18-71-10061 from the Russian Science Foundation.
The work of Y.A.B. was finished within the Project Teams framework of MIEM HSE.

\section*{Author contribution statement}
Y.A.B. and L.Y.B. developed the model.
S.V.Z. and A.L.K. performed the simulations and the data analysis. 
All the authors were involved in preparing the manuscript. 
All the authors have read and approved the final manuscript.

\appendix
\section{Basic equations in the case of axial symmetry}
\label{appendix-axial}
In this appendix we will write the basic equations in cylindrical 
coordinates $r,z$ taking into account the axial symmetry. 

Eq. (\ref{Poisson}) in $(r, z)$ coordinates takes the form
\begin{equation}\label{eq1rz}
- \bigg(\dfrac{\partial \phi}{\partial r} \dfrac{\partial \varepsilon}{\partial r} + \dfrac{\partial \phi}{\partial z} \dfrac{\partial \varepsilon}{\partial z} + \varepsilon \bigg(\dfrac{\partial^2 \phi}{\partial r^2} + \dfrac{1}{r} \dfrac{\partial \phi}{\partial r} + \dfrac{\partial^2 \phi}{\partial z^2}\bigg)\bigg) = \frac{\rho(r, z, t)}{\varepsilon_{0}}.
\end{equation}

The Nernst--Planck equation (\ref{Fokker}) in the $(r, z)$ coordinates can be written as follows:
\begin{eqnarray}
\dfrac{\partial n_c}{\partial t} = \dfrac{q_c}{\gamma_c} \bigg(\dfrac{\partial \phi}{\partial r} \dfrac{\partial n_c}{\partial r} + \dfrac{\partial \phi}{\partial z} \dfrac{\partial n_c}{\partial z} + n_c \bigg(\dfrac{\partial^2 \psi}{\partial r^2} + \dfrac{1}{r} \dfrac{\partial \phi}{\partial r} + \dfrac{\partial^2 \phi}{\partial z^2}\bigg)\bigg) + \dfrac{\partial n_c}{\partial r} \dfrac{\partial D_c}{\partial r} + \dfrac{\partial n_c}{\partial z} \dfrac{\partial D_c}{\partial z} +\nonumber\\ D_c \bigg(\dfrac{\partial^2 n_c}{\partial r^2} + \dfrac{1}{r} \dfrac{\partial n_c}{\partial r} + \dfrac{\partial^2 n_c}{\partial z^2}\bigg) - \textbf{v}\cdot\nabla n_c-n_c \nabla\cdot\textbf{v}.
\end{eqnarray}

In the same way, we rewrite eq.(\ref{Fokker_2}) for concentrations of ions, $n_i (r, z, t)$ ($i=\pm$),
\begin{eqnarray}
\dfrac{\partial n_i}{\partial t} = \dfrac{q_i}{\gamma_i} \bigg(\dfrac{\partial \phi}{\partial r} \dfrac{\partial n_i}{\partial r} + \dfrac{\partial \phi}{\partial z} \dfrac{\partial n_i}{\partial z} + n_i \bigg(\dfrac{\partial^2 \phi}{\partial r^2} + \dfrac{1}{r} \dfrac{\partial \phi}{\partial r} + \dfrac{\partial^2 \phi}{\partial z^2}\bigg)\bigg) + \dfrac{\partial n_i}{\partial r} \dfrac{\partial D_i}{\partial r} + \dfrac{\partial n_i}{\partial z} \dfrac{\partial D_i}{\partial z} +\nonumber\\ D_i \bigg(\dfrac{\partial^2 n_i}{\partial r^2} + \dfrac{1}{r} \dfrac{\partial n_i}{\partial r} + \dfrac{\partial^2 n_i}{\partial z^2}\bigg) - \textbf{v}\cdot\nabla n_i-n_i \nabla\cdot\textbf{v}.
\end{eqnarray}
In our simulation we deal with an incompressible liquid, so that $\nabla\cdot\mathbf{v}=0$.

The boundary conditions for the electrostatic potential have the following form
\begin{equation}
\phi_{d}^{(sub)}=\phi_{sub}=\Phi,
\end{equation}
\begin{equation}
\varepsilon \bigg(r \dfrac{\partial \phi_{d}^{(+)}}{\partial r} + (z+R\cot{\theta}) \dfrac{\partial \phi_{d}^{(+)}}{\partial z}\bigg) = r \dfrac{\partial \phi_{d}^{(-)}}{\partial r} + (z+R\cot{\theta}) \dfrac{\partial \phi_{d}^{(-)}}{\partial z},
\end{equation}
\begin{equation}
\phi_d^{(-)} = \phi_d^{(+)}, \quad \phi \big| _{|r| \to \infty} = 0.
\end{equation}
where $R$ is the radius of the substrate and $\theta$ is the contact angle.
The boundary conditions for the concentrations take the following form
\begin{equation}
r \dfrac{\partial n_{c,\pm}}{\partial r} + (z+R\cot{\theta}) \dfrac{\partial n_{c,\pm}}{\partial z} + \dfrac{q_{c,\pm}n_{c,\pm}}{k_{B} T} \bigg(r \dfrac{\partial \phi_{d}^{(+)}}{\partial r} + (z+R\cot{\theta}) \dfrac{\partial \phi_{d}^{(+)}}{\partial z}\bigg) = 0,
\end{equation}
\begin{equation}
\dfrac{\partial n_{c,\pm}}{\partial z} + \dfrac{q_{c,\pm}n_{c,\pm}}{k_{B} T}  \dfrac{\partial \phi_{d}^{(sub)}}{\partial z} = 0.
\end{equation}

\section{Temperature distribution inside the droplet}
\label{appendix-temperature}
The field of temperature, $T(\mathbf{r},t)$, inside the droplet can be obtained from the numerical solution of the heat transfer equation taking into account the liquid motion
\begin{equation}
\dfrac{\partial T}{\partial t} + \textbf{ v} \cdot \nabla T = \kappa \Delta T,
\end{equation}
where $\kappa$ is the thermal diffusivity of the liquid. The boundary conditions for this equation have the following form $\partial T / \partial r =  0$ at $r = 0$; $T = T_0$ at $z = 0$; $\partial T / \partial {\mathbf{n}} = -Q_0(r)/k$ at the droplet free boundary, where $k$ is the thermal conductivity of the liquid; $Q_0(r) = LJ_s(r)$ is the heat flow with the specific heat of vaporization, $L$, and the vaporization flux density, $J_s(r)$, which can be calculated by the following interpolation formula~\citep{Deegan2000}
\begin{equation}\label{Jr}
J_s (r) = J_0(\theta) (1-r^2/R^2)^{-(1/2-\theta/\pi)},
\end{equation}
where $J_0 (\theta)$ can be determined by the following expressions~\citep{HuLarson2002}:
\begin{eqnarray}
J_0 (\theta) / (1-\Lambda(\theta)) &=& J_0(\pi/2) (0.27 \theta^2 + 1.3),\\
\Lambda(\theta) &=& 0.2239(\theta-\pi/4)^2+0.3619,\\
J_0(\pi/2) &=& D u_s/R.
\label{J0}
\end{eqnarray}
We note that in our case it is not necessary to solve the heat transfer equation inside the substrate, because we assume that the substrate connects with a thermostat, fixing its temperature $T_0$.

Note that the droplet surface of the spherical segment shape can be parameterized by the following equations
\begin{equation}
h(r,t)=\frac{R(\cos\varphi(r,t)-\cos\theta(t))}{\sin\theta(t)};\qquad
\varphi(r,t)=\arcsin\left(\frac{r\sin\theta(t)}{R}\right),
\label{hFromR}
\end{equation}
where $\varphi$ is the angle between the normal vector and the vertical axis $z$.

\section{Velocity distribution inside the droplet for the Marangoni flux case}
\label{appendix-velocity}
In the axial symmetry case, the Navier--Stokes equations can be rewritten in terms of vorticity, $\gamma(r,z)={\p u_r}/{\p z}-{\p u_z}/{\p r}$, 
and stream function, $\psi$, as follows \cite{Barash2009}:
\begin{eqnarray}
\frac{\p}{\p t}\gamma(r,z)+(\vvect\cdot\mynabla)\gamma(r,z)&=&
\nu \left(\Delta\gamma(r,z)-\frac{\gamma(r,z)}{r^2}\right),\\
\frac{\p^2\psi}{\p r^2}-\frac1{r}\frac{\p\psi}{\p r}+
\frac{\p^2\psi}{\p z^2}&=&r\gamma,
\end{eqnarray}
where the stream function is related to the velocity components {\sl via} the following relations
${\p\psi}/{\p z}=ru_r$, ${\p\psi}/{\p r}=-ru_z$.

The boundary conditions: $\gamma=0$ at $r=0$, $\gamma=\p u_r/\p z$ at $z=0$;
$\gamma=\eta^{-1}{d\sigma}/ds+2u_\tau d\varphi/ds$
at the droplet boundary; $\psi=0$ at all the boundaries: droplet boundary, axis of symmetry ($r=0$) and substrate-fluid interface ($z=0$).
Here ${d\sigma}/{ds}=\sigma^{\prime}\p T/\p s$ is the derivative of the surface tension with respect to the arc length along the droplet surface, where the distribution of temperature discussed above is taken into account; $\sigma^{\prime}=\p \sigma/\p T$ is derivative of surface tension with respect to the temperature.
Variable $\varphi$ is the angle between the normal vector and the vertical axis $z$, so that
$\p T/\p s = \cos\varphi\cdot \p T/\p r - \sin\varphi\cdot \p T/\p z$, and the tangent component of velocity is $u_\tau = u_r \cos\varphi-u_z \sin\varphi$.
For the spherical shape of the droplet, described by equations~(\ref{hFromR}),  
the arc length $s$ is described by $\varphi$ {\sl via} the relation $s = R\varphi / \sin\theta$.
In particular, at the contact line $s_{max} = R\theta / \sin\theta$.

\section{Boundary conditions for Navier--Stockes equations for the capillary fluid flow case}
\label{appendix-capillary}

The boundary conditions for Navier--Stockes equations discussed in Appendix~\ref{appendix-velocity} were derived with the assumption that $u_n = 0$. That is why the capillary flow was not taken into account in this derivation. In order to take into account the capillary fluid flow, it is necessary to rewrite the boundary conditions in a more general form. For the vorticity at the droplet boundary we have
\begin{equation}
\gamma = \frac1\eta \frac{d\sigma}{ds} + 2u_\tau\frac{d\varphi}{ds}-2\frac{\p u_n}{\p s},
\label{bound_gamma}
\end{equation}
where $u_n = u_r \sin\varphi + u_z \cos\varphi = -\p \psi / (r\p s)$.
For the stream function $\psi=0$ at the axis of the droplet symmetry and the droplet-substrate interface. However, at the droplet boundary we have
\begin{equation}
\psi = \int_0^s (-r u_n) ds.
\label{bound_psi}
\end{equation}

As it follows from the mass conservation law in an infinitely small cylindrical volume near the droplet surface, on the droplet surface the following relation
\begin{equation}
u_n - \frac{\p h(r,t)}{\p t}\cos\varphi = \frac{J_s(r,t)}{n}
\label{we_find_u_n}
\end{equation}
is fulfilled, where $n$ is the liquid density.

Note that
$m = n\pi h_0^2 (R-h_0/3)$, where $h_0 = R(1-\cos\theta)/\sin\theta$, i.e.
$m = \frac13\pi R^3n (3-\tan(\theta/2))\tan^2(\theta/2)$. Therefore,
\begin{equation}
\frac{dm}{dt} = -\frac12n\pi R^3 \frac{\tan\frac{\theta}{2}-2}{\cos^2\frac{\theta}{2}}\tan\frac{\theta}{2}\cdot\frac{d\theta}{dt}.
\end{equation}
Thus, using the relation 
\begin{equation}
\frac{dm}{dt}=-\int_0^{s_{max}} 2\pi r(s) J(r(s)) ds,
\end{equation}
one can calculate $d\theta/dt$. Further, using equations~(\ref{hFromR}), we get
\begin{equation}
\frac{\p h(r,t)}{\p t} =
\frac{R }{\sin^2\theta}\left(1-\frac{\cos\theta}{\cos\varphi}\right)\frac{d\theta}{dt}.
\end{equation}
Knowing $\p h(r,t)/\p t$, we obtain $u_n$ on the droplet boundary using eq.~(\ref{we_find_u_n}).
Knowing $u_n$, we can use boundary conditions (\ref{bound_gamma}) and (\ref{bound_psi}).
\bibliography{refs}
\end{document}